\documentclass{ws-procs961x669}            

\def \bea {\begin{eqnarray}}
\def \eea {\end{eqnarray}}

\begin{document}
\title{Reliable Equations of State of Viscous Strong and Electroweak Matter}
\author{A. Tawfik$^*$}
\address{Egyptian Center for Theoretical Physics (ECTP), Cairo, Egypt\\
World Laboratory for Cosmology And Particle Physics (WLCAPP), Cairo, Egypt \\
$^*$E-mail: a.tawfik@cern.ch}

\begin{abstract}
For the first time, a reliable estimation for the equations of state (EoS), bulk viscosity, and relaxation time, at temperatures ranging from a few MeV up to TeV or energy density up to $10^{16}~$GeV/fm$^3$. This genuine study covers both strong and electroweak epochs of the early Universe. Non--perturbation (up, down, strange, charm, and bottom quark flavor) and perturbative calculations (up, down, strange, charm, bottom, and top quark flavors), are phenomenologically combined, at vanishing baryon--chemical potential. In these results, calculations from Polyakov linear--sigma model (PLSM) of the vacuum and thermal condensations of the gluons and the quarks (up, down, strange, and charm flavors) are also integrated. Furthermore, additional degrees of freedom (photons, neutrinos, charged leptons, electroweak particles, and scalar Higgs boson) are found significant along the entire range of temperatures. As never done before, the present study brings the standard model of elementary particles closer to the standard model for cosmology.

\end{abstract}


\bodymatter

\section{Introduction}

Understanding the strong and electroweak epochs of the early Universe, at least, through determining their equations of state (EoS) is impactful for various cosmological studies, such as the nucleosynthesis and the cosmological large--scale structure. Until present date, cosmology is dictated by the standard model of cosmology (SMC). SMC assumes that the cosmic background is isotropically and homogeneously characterized by an ideal fluid. With this regard, we recall that a first proposal that the viscous coefficients are also connected with particle physics was presented ref. \cite{misner1968isotropy,Zeldovich:1983cr}. It was also assumed that the influence of viscous coefficients becomes significant first at the end of the lepton era \cite{husdal2016viscosity}, i.e., during neutrino decoupling epoch or temperature $T\simeq 10^{10}~$K ($\simeq 1~$MeV) which takes place not long after the end of the QCD phase transition from colored quark--gluon plasma (QGP) to colorless hadrons. The QCD phase transition \cite{Tawfik:2004sw,Tawfik:2010pm,STAR:2015vvs,Hakk:2021aav} takes place at $T\simeq 160~$MeV. Recent studies concluded that the viscous coefficients likely impact early epoches of the Universe \cite{Tawfik:2019jsa}. 

Taking advantage of recent developments in ultra--relativistic nuclear experiments and non--perturbative lattice QCD simulations, and perturbative calculations, various thermodynamic quantities including pressure $p$, energy density $\rho$, bulk viscosity $\zeta$, and relaxation time $\tau$ are calculated up to the TeV-scale and a reliable evolution of the early Universe could be achieved. The present study offers an access to deep epochs of the early Universe with $T$ up to TeV-scale and $\rho$ up to $10^{16}~$GeV/fm$^3$.

For Friedmann--Lemaitre--Robertson--Walker (FLRW) metric 
\begin{equation}  
ds^{2}=dt^{2}-a(t)^{2} \left[\frac{dr^{2}}{1-k r^2}+r^{2}\left(d\theta^{2}+\sin^{2}\theta d\phi^{2}\right) \right], \label{1}
\end{equation}
where $a(t)$ is the dimensionless scale factor, $k=\{-1,0,+1\}$ represents elliptical, flat (Euclidean), and hyperbolic cosmic space, or negative, flat, and positive curvature, respectively. The theory of general relativity (GR) inters the play, when temporal evolution of the line element $s(t)$ can be determined. To this end, GR has to be combined with the matter--energy content of the cosmic geometry,  
\begin{equation}
R_{\mu \nu}-\frac{1}{2}g_{\mu \nu}\, R + \Lambda_{\mu \nu}=\frac{8 \pi}{3}\, T_{\mu \nu},  \label{ein}
\end{equation}
where $\mu$, $\nu$ run over $0$, $1$, $2$, and $3$. 

Then, the Einstein field equations in natural units read
\begin{eqnarray}    
H(t)^{2} &=& \frac{8 \pi}{3} \;\rho(t) - \frac{k}{a(t)^2} + \frac{\Lambda}{3}, \label{dH}\\
\dot H(t) + H(t)^2 &=& -\frac{4 \pi}{3} \; \left[\rho(t) + 3p_{\mathtt{eff}(t)} \right]  + \frac{\Lambda}{3}, \label{drho}
\end{eqnarray}
where $H(t)=\dot a(t)/a(t)$ is the Hubble parameter. From (\ref{dH}) and (\ref{drho}), the time evolution of the Hubble parameter can be deduced 
\bea
\dot H(t) &=& -4 \pi\, \left[\rho(t) + p_{\mathtt{eff}}(t)\right] + \frac{k}{a(t)^2}. \label{Eck-Hh0}
\eea
An EoS relating $p$ with $\rho$ is needed to have a closed system of equations.

The present script is organized as follows. The most reliable non--perturbative and perturbative calculations are reviewed in section \ref{sec:LQCDsims}. Our results on EoS, bulk viscosity and relaxation time for almost the entire SM dof are presented in section 
\ref{sec:EoSZetaTau}. The conclusions are outlined in section \ref{sec:cncl}.

\section{Most reliable non-perturbative and perturbative calculations}
\label{sec:LQCDsims}

For relaible EoS, various general considerations have been assumed in the recent non-perturbative lattice QCD simulations \cite{Borsanyi:2016ksw}. The first one is the quark masses. For a dynamical dependence of the quark masse $m_q$ and the lattice spacing $a$ on the guage coupling $\beta$, four flavor staggered action with $4$ levels of stout smearing has been utilized, where $u$, $d$, $s$, and $c$ quarks are realized as $2+1+1$ and $3+1$ \cite{Borsanyi:2016ksw}, i.e., except for strange and charm quark masses, $m_{ud}=R\times m_s^{st}(\beta)$, with $m_s^{st}(\beta)$, $1/R=27.63$ and $\beta$ is the guage coupling. The $c$ quark mass is given as a function of the gauge coupling; $m_c=C\times m_s^{st}(\beta)$, with $C=11.85$. Although the degenerate masses of the light quarks, a small isospin asymmetry could also be included, analytically \cite{Borsanyi:2016ksw}. 

The second general consideration is the temperature $T$, which can primarily be determined as a function of the temporal lattice dimension; $T=(a N_{\tau})^{-1}$. Alternatively, varying the gauge coupling $\beta$ leads to changing $T$, as well, even if the spacial and temporal lattice dimensions might not be varied. The gauge coupling cann't only allow for varying $T$, but also it measures the pseudoscalar pion mass $m_{\pi}$ and the Wilson--flow based scale $\omega_0$, where $\omega_0=0.153\pm0.001~$fm and $m_{\pi}=712\pm5~$MeV. At $T=0$, $\omega_0$ gives the inverse flow time \cite{Borsanyi:2012zs}.

The third general consideration is the lattice QCD computational procedure. After applying a Wilson--flow equation, the clover definition of the topological charge was made in $2+1+1$ and $3+1$ ensembles. To make the computational process more economic, an adaptive step size integration scheme was utilized. The time flow $(8 T^2)^{-1}$ was introduced to estimate the finite $T$ of both ensembles, where a variation in the time flow is also allowed. This procedure has greatly contributed to reducing the systematic errors. To control their simulations, it should be checked whether these configurations lead to saturated susceptibility, at large flow times \cite{Borsanyi:2016ksw}. To determine the topological susceptibility, topological charge was utilized with and without rounding.

The fourth general consideration is the acceptance accuracy. As already outlined for the $2+1+1$ simulations, same configurations have been used also for $3+1$ simulations. Here, the mass ratio of the charm quark and the degenerate lighter quarks ($u$, $d$, and $s$) was taken to be $11.85$. As mentioned, for the masses of up-- and down--quarks, the physical values are used, while the mass of the strange quark is a function of the guage coupling, $m_s^{\mathtt{st}}(\beta)$. At $T=0$, simulations were done on $64 \times 32^3$ lattice with seven values of the lattice spacing descendingly ranging from $0.15$ to $0.06~$fm. But at finite $T$, the same parameters as in $2+1+1$ case were used. The topological charge is measured for every Hybrid Monte--Carlo trajectory. The configurations leading to a topology change are rejected. In other words, configurations are generated, at fixed topology. Quantitatively, this leads to an acceptance probability of about $40\%$ for the coarsest lattice, but higher probabilities for the finest ones.   

In the following section we review how various thermodynamic quantities have been determined.

\subsection{Lattice QCD Equation of State in Non--Perturbative Regime}

The inclusion of up--, down--, strange--, and charm--quark in non--perturbative lattice QCD simulations \cite{Cheng:2007wu,Levkova:2009gq} assumes that the masses of some quarks are very heavy (partially quenched). Other lattice QCD simulations assume that the four quarks are non--degenerate with unphysical masses \cite{Burger:2013hia,MILC:2013ops}. The ultimate goal is to carry out non--perturbative simulations with dynamical quarks and physical masses \cite{Borsanyi:2011wyg}. To this end, $2+1+1$ simulations with staggered action, and $4$ levels of stout smearing have been carried out in ref. \cite{Bellwied:2015lba}. 

One restriction in the lattice QCD simulations is the universal assumption of thermal equilibrium. This might not fully true, especially because the hadron and parton matter is undergoing phase transition, whether prompt of slows crossover, at critical temperatures, i.e., non--equilibrium due to changes in the underlying dynamics, symmetry, and degrees of freedom. Another restriction is that the temperature dependence of pressure $p$, energy density $\rho$, and entropy $s$ can be deduced from the trace anomaly, at vanishing chemical potential,  
\bea
\frac{I(T)}{T^4} &=& \frac{\rho - 3 p}{T^4}, \label{eq:IT1} \\
\frac{\rho + p}{T^4} &=& \frac{s}{T^3}.  \label{eq:TH1}
\eea
A third restriction is the common practice to get rid of the temperature independent divergence of the trace anomaly so that its physical value can be evaluated, accurately. A vanishing--$T$ ensemble is subtracted from each finite--$T$ ensemble. As this method doesn't work well at high temperatures, where short lattice spacing and increasing autocorrelation times and computational costs are likely, a renormalization ensemble is generated for each finite--$T$ ensemble, at exactly half of its temperature \cite{Borsanyi:2016ksw}. The physical trace anomaly in a wide range of temperatures including low $T$, can be obtained from the subtraction of each half--$T$ from its finite--$T$ ensemble, $[I(T)-I(T/2)]/T^4$. The resulting $\rho$ and $s$ can straightforwardly be estimated from Eqs. (\ref{eq:IT1}) and (\ref{eq:TH1}). 

Up to four quark flavors were included in the non--perturbative lattice QCD simulations. This was achieved by several steps of tree--level corrections.

\subsection{Inclusion of $c$--quark in Non--Perturbative Lattice QCD simulations}
\label{sec:lqcdpert}

It was concluded \cite{Andersen:2010wu} that the free energy calculated in next--to--next--to--leading order (NNLO) Hard Thermal Loop (HTL) perturbation theory is in good agreement with $2+1$ and $2+1+1$ non--perturbative lattice QCD simulations, where $c$--quark is taken massless in the perturbative calculations but assigned the physical mass in the non--perturbative simulations. The mass of $c$--quark was perturbatively \cite{Laine:2006cp} and non--perturbatively \cite{Borsanyi:2016ksw} estimated. In the perturbative calculations, the effect of heavy quarks is determined to a lower leading order, and accordingly, it was concluded that this refers to $(3+1)$ quark flavors pressure normalized to $(3)$ quark flavors pressure $p$. On the other hand, when comparing $p$ with and without $c$--quark in both non--perturbative and perturbative calculations, an excellent agreement ($<3\%$) was obtained \cite{Borsanyi:2016ksw}. The tree--level correction due to $c$--quark reads
\bea
\frac{p^{(2+1+1)}(T)}{p^{(2+1)}(T)} &=& \frac{SB^{(3)}(T)+F_Q(m_c/T)}{SB^{(3)}(T)}, \label{eq:chrm1}
\eea
where $SB$ stand for Stefan--Boltzmann approximation and $F_Q(m_c/T)$ is the free energy density of a free quark with mass $m_c=1.29~$GeV.

The following section elaborates how the $b$--quark flavor is included non--perturbative lattice QCD simulations. This was achieved by several steps of tree--level corrections.

\subsection{Inclusion of $b$--quark in Non--Perturbative Lattice QCD Simulations}
\label{sec:lqcdpert}

The success with the inclusion of the $c$--quark encouraged a recent attempt with the bottom quark, especially that the $2+1+1$ non--perturbative simulations up to $T\lesssim 1~$GeV  control the accuracy of the proposed procedure and the perturbative contributions likely dominate, at $T\gtrsim 500~$MeV) \cite{Kajantie:2002wa,Brambilla:2006wp}. Recent perturbative calculations were performed up to $\mathcal{O}(g^6 \log g)$ \cite{Borsanyi:2016ksw}. Thus, the inclusion of $b$--quark and a continuation to higher $T$ look straightforward. A tree--level correction for the $b$--quark similar to that for the $c$--quark, Eq. (\ref{eq:chrm1}), was suggested. 

It was concluded that tree--level correction for $b$--quark works well. When comparing the ratio of the massless $2+1+1$ to $2+1+1+1$ perturbative pressure with the pressure ratio in the SB limit, an excellently agreement ($<0.3\%$) is obtained. With a phenomenological approach similar to Eq. (\ref{eq:chrm1}), reliable non--perturbative lattice QCD simulations become feasible, at $500~$MeV $<T<10~$GeV. This $T$--range apparently covers various epochs in the early Universe, where different phase transitions and accordingly different dynamics and degrees of freedom become dominant. 

In the following section, we review features of the perturbative calculations up to TeV temperatures.

\subsection{Perturbative Calculations up to TeV Temperature--Scales}
\label{sec:pQCDsims}

In the previous sections, we have discussed on the various restrictions of the non--perturbative lattice QCD simulations and their possible extensions to a large number of quark flavors and to higher temperatures. The perturbative calculations, on the other hand, allow to cover much higher temperatures and to include more quark flavors. By combining recent perturbation calculations up to a largest leading order with the non--perturbative lattice QCD simulations, relaible EoS for strong and electroweak matter can be deduced \cite{Laine:2015kra,Laine:2013raa}. It is conjectured that this covers temperatures up to $\sim 200~$TeV.

\subsubsection{QCD Domain}

In this domain, $0.2\lesssim T\lesssim 1~$GeV, the gluons and the lightest four quark flavors and the gluons are partonic dof of the strongly interacting matter \cite{Laine:2015kra}. The perturbative corrections to the ideal (masslss, noninteracting) Stefan-Boltzmana (SB) EoS can be determined up to different orders of strong coupling constant $\mathcal{O}(g)$. Although, the perturbative contributions up to $\mathcal{O}(g^6 \log g)$ are well known \cite{Linde:1980ts,Gross:1980br,Kajantie:2002wa}, where $g$ is expressed as function of $N_c$ the colors dof, $N_f$ the massless quark flavors dof, and $\mu_f$ the quark chemical potential, only $\mathcal{O}(g^2)$ terms, at $T=0$ but $\mu_f\neq 0$, have been precisely, analyzed. 

Besides the contributions of gluons up to $\mathcal{O}(g^6 \log g)$ and that of NLO $\mathcal{O}(g^2)$, it was found that these are similar to that of LO $\mathcal{O}(g^0)$, where no $\mathcal{O}(g^6 \log g)$ calculations for finite quark masses are available so far \cite{Laine:2006cp}. An alternative procedure was suggested. This starts with $N_f=0$ and $N_c=3$ corresponding to very heavy quark flavors. Then, it calculates the change in the pressure by lowering the quark masses down to their physical values\footnote{Decreasing the quark masses increases the thermodynamic pressure.}. This procedure suggests that the perturbative calculations are based on the grand--canonical pressure (or free energy) \cite{Borsanyi:2016ksw}. As discussed earlier, the non--perturbative lattice QCD simulations starts with the trace anomaly (also known as the interaction measure) $I(T,\mu_f,\cdots)$, Eq. (\ref{eq:IT1}), from which the various thermodynamic quantities can be determined. Based on this renormalization procedure, the ultraviolet divergences are likely removed. 

In order to extend the perturbative calculations to the electroweak domain, the so--called {\it hard modes} should first be removed through either integration with respect to momenta or summation over the Matsubara frequencies, $2 \pi T$, $g T$, $g^2 T$, $\cdots$. Second, the effective mass parameters and the gauge couplings have to be specified and normalized in $\overline{\mathtt{MS}}$--scheme. Third, the connection factors should then be estimated, at changing $T$ and fixed $\Lambda_{\overline{\mathtt{MS}}}$. These can be achieved, when multiplying the non--perturbative lattice results for $N_f=0$, $N_c=3$, finite quark masses by the corresponding ones obtained in the SB limit, Eq. (\ref{eq:chrm1}). The connection factors facilitate the inclusion of heavier quarks, at temperatures greater that strong QCD scale parameter $\Lambda_{QCD}\sim200~$MeV.

\subsubsection{Electroweak Domain}

As discussed, the perturbative calculations are initiated from the free energy (or thermodynamic pressure), the Higgs potential parameters $v^2(\bar{\mu})$ and $\lambda(\bar{\mu})$ are given as functions of the normalization scale $\bar{\mu}$. Assuming that the bottom and top quarks weakly interact, the free energy, at finite $T$ and $\bm{\mu}$, reads \cite{Laine:2006cp}
\bea
f(v, T, \bm{\mu}) &=& -\frac{1}{2} \nu^2(\bar{\mu}) v^2 + \frac{1}{4} \lambda(\bar{\mu}) v^4 + \sum_i^{n_q+n_g} \pm {J}_i (m_i(v),T,\mu_i), \label{eq:EWfenergy}
\eea  
where $n_q\; (n_g)$ are the number of quarks (gluons), $v$ is the Higgs expectation value, $m_i(v)$ is the tree--level mass of $i$--th particle, $\pm$ stand for bosons and fermions, respectively, and ${J}_i (m_i(v),T,\mu_i)$ counts for the contributions of the physical dof (scalars, vectors, and fermions). The normalization scale $\bar{\mu}$ could be fixed to $100~$GeV\footnote{This is a model--depending assumption. Originally, the electroweak theory has different scales along its wide range of temperatures and chemical potentials.}. Nevertheless, the electroweak free energy, Eq. (\ref{eq:EWfenergy}), can be determined, at temperatures far beyond the electroweak scale $\bar{\mu}\sim100~$GeV. 

As the proposed EoS characterizes electroweak matter, a few remarks on the nature of the phase transitions is now in order. In some perturbative calculations, it was concluded that the electroweak EoS seems approaching that of an ideal gas \cite{Laine:2006cp}, while in others significant deviations have been obtained \cite{Gynther:2005dj}. With this regard, it should be considered that the strength of electroweak phase transition is determined by different SM Lagrangian parameters, which so--far aren't determined. The recently estimation of the Higgs mass allowed for a smooth crossover and electroweak baryogenesis \cite{Kajantie:1996mn,Aoki:1999fi}.

\subsection{Combining non--perturbative with perturbative EoS}
\label{sec:rslts}

\begin{figure}[htb!]
\centering{ 
\includegraphics[width=4cm,angle=-90]{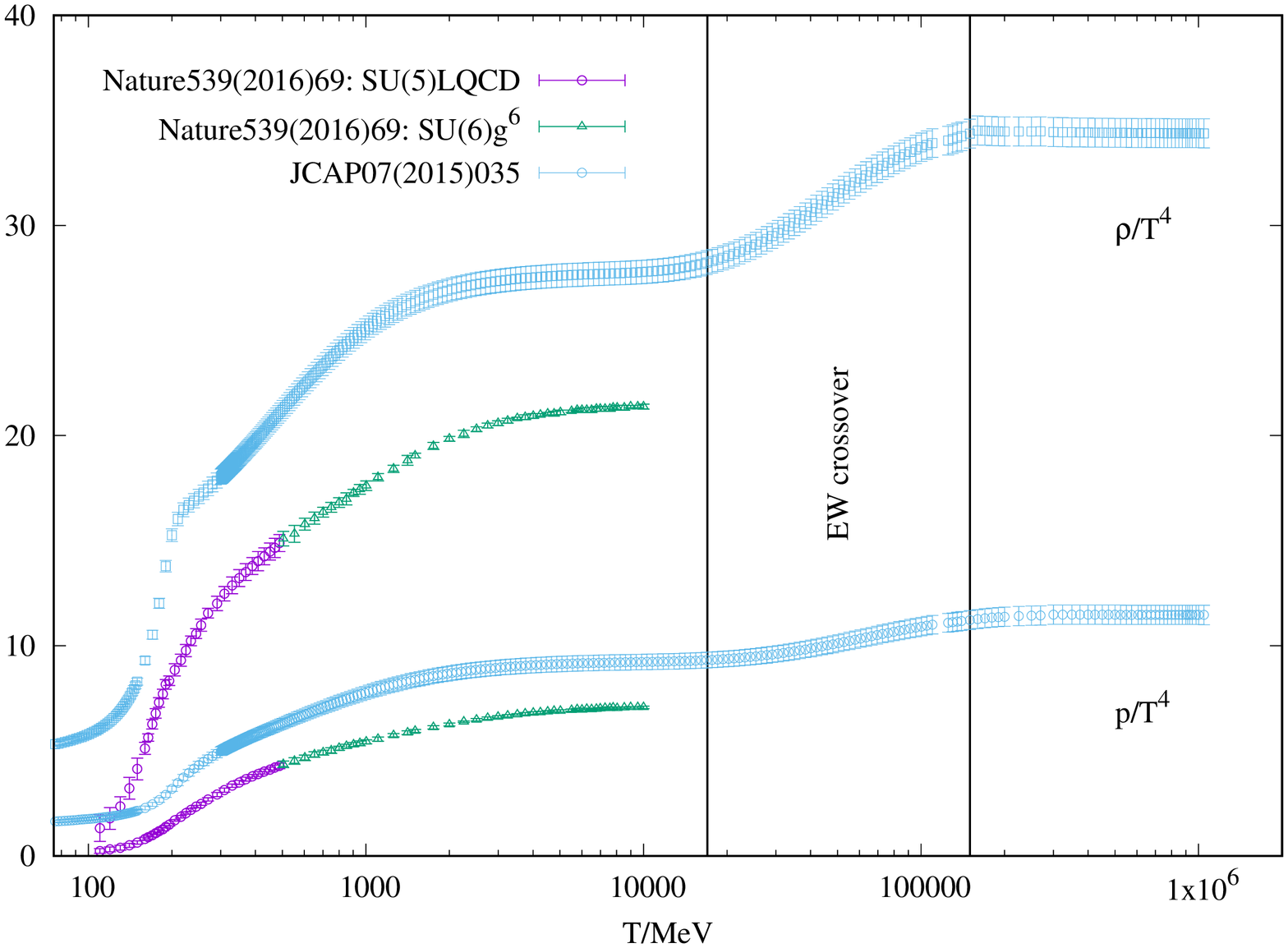} 
\includegraphics[width=4cm,angle=-90]{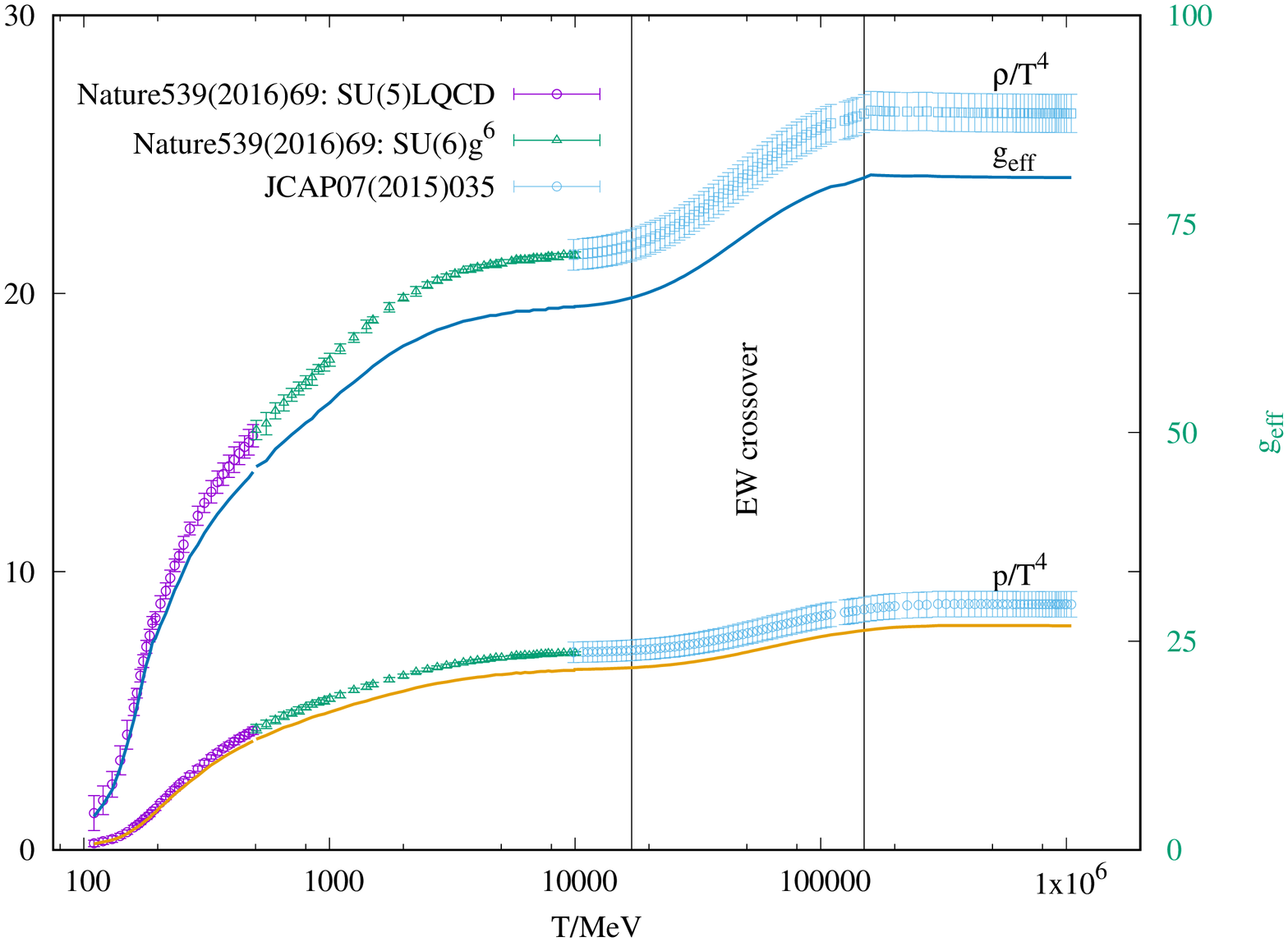} 
\caption{Left panel: a comparison between non--perturbative lattice QCD simulations \cite{Borsanyi:2016ksw} and the perturbative calculations \cite{Laine:2006cp} for normalized pressure and energy density as functions of temperatures. Right panel: the same as in the left panel but here the perturbative calculation are rescalled (see main text). The vertical lines approximately determine the temperatures, where the electroweak crossover takes place. The curves in right panel present the effective dof $\mathtt{g_{eff}}(T)$ related to pressure and energy density. }
\label{fig:x}
}
\end{figure}

As discussed, the proposal to combine reliable non--perturbative with perturbative calculations was already applied in various studies. In the present study, we combine the non--perturbative \cite{Borsanyi:2016ksw} with the perturbative calculations \cite{Laine:2006cp} for different thermodynamic quantities. 

The temperature dependence of the normalized pressure and energy density calculated non--perturbatively and perturbatively is illustrated in the left panel of Fig. \ref{fig:x}. Here, the temperature covers up to $\sim 10~$GeV and $\sim 1~$TeV in non--perturbative and perturbative simulations, respectively. It is obvious that although both types of simulations look quantitatively different, they are qualitatively similar. We find that the perturbative results are significantly larger than the non--perturbative ones. 

There are many reasons supporting the assumption that the non--perturbative lattice QCD simulations are more reliable than the perturbative ones, especially at low temperatures. In this range of temperatures, even the vacuum energy change is best evaluated by non--perturbative simulations. On one hand, the non--perturbative lattice QCD simulations are most reliable, at low temperatures, the so--far perturbative calculations are not as accurate even at high temperatures. Thus, we concretely propose that the non--perturbative lattice QCD simulations are most reliable, at $T\gtrsim\Lambda_{\mathtt{QCD}}\approx 200~$MeV up to a few GeV \cite{Brodsky:2002nb}, while the perturbative calculations are the only approach possible, at temperatures up to the TeV--scale. From Fig. \ref{fig:x}, a temperature--indenedent systematic difference of $\sim20\%$ between both sets of calculations\footnote{It was reported in Ref. \cite{Borsanyi:2016ksw} that this difference  reads $7 - 17\%$ and is temperature dependent!} is obtained. Also, we assumed that the systematic difference of $\sim20\%$ ins the same for both pressure and energy density.  

For a contineous temperature--dependence of pressure and energy density, we propose to rescale the perturbative calcuations, at high temperatures \cite{Laine:2006cp}. The rescaling factor is phenomenologically adjusted to match the perturbative calculations, at high temperatures, with the non--perturbative lattice QCD simulations, at low temperatures. We assume that a rescaling factor of $0.77$ remains constant, at the entire range of temperatures, right panel of Fig. \ref{fig:x}. The non--perturbative lattice QCD simulations aren't rescaled, at all. In both thermodynamic quantities, both non--perturbative ($T$ up to $\sim 10~$GeV) and perturbative ($10~\mathtt{GeV}\lesssim T\lesssim 1~$TeV) calculations are phenomenologically combined. The effective dof $\mathtt{g_{eff}}(T)$ corresponding to pressure, $\mathtt{g_{eff}}(T)=p(T)/p_0$, and to energy density, $\mathtt{g_{eff}}(T)=\rho(T)/\rho_0$, are also depicted in the right panel of Fig. \ref{fig:x}, where $\rho_0=(\pi^2/30)T^4$ is the energy density and $p_0$ is the pressure for an ideal gas of scalar massless bosons $p_0=(\pi^2/90)T^4$. The perfect matching of $\mathtt{g_{eff}}(T)$, at low and high temperatures, supports the conclusion that the temperature--independent proposed rescaling, $0.77$, seems precise.

\section{EoS, Bulk viscosity and Relaxation time for almost entire SM dof}
\label{sec:EoSZetaTau}
\subsection{EoS for Strong and Electroweak Matter}

\begin{figure}[htb!]
\centering{
\includegraphics[width=4cm,angle=-90]{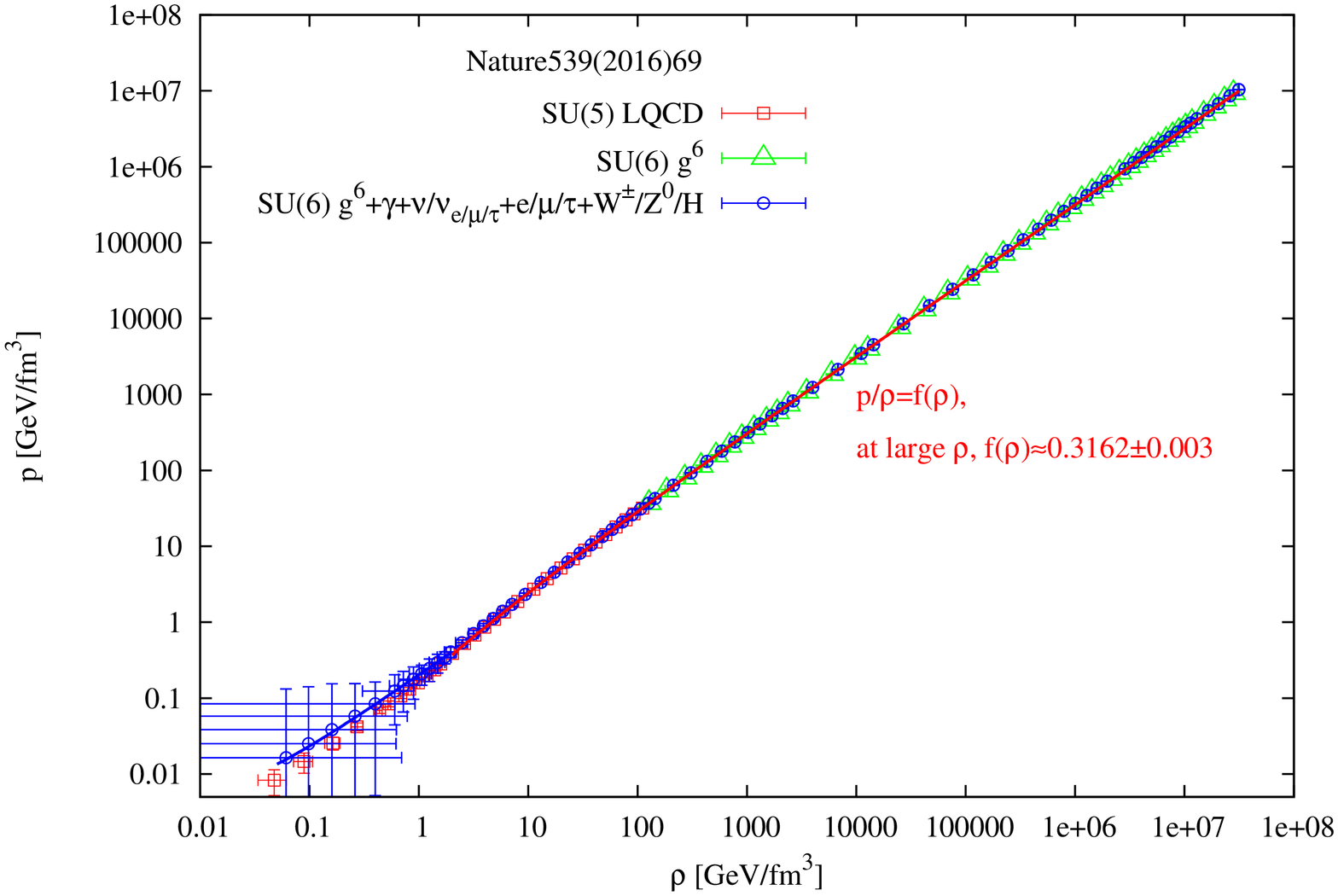}
\includegraphics[width=4cm,angle=-90]{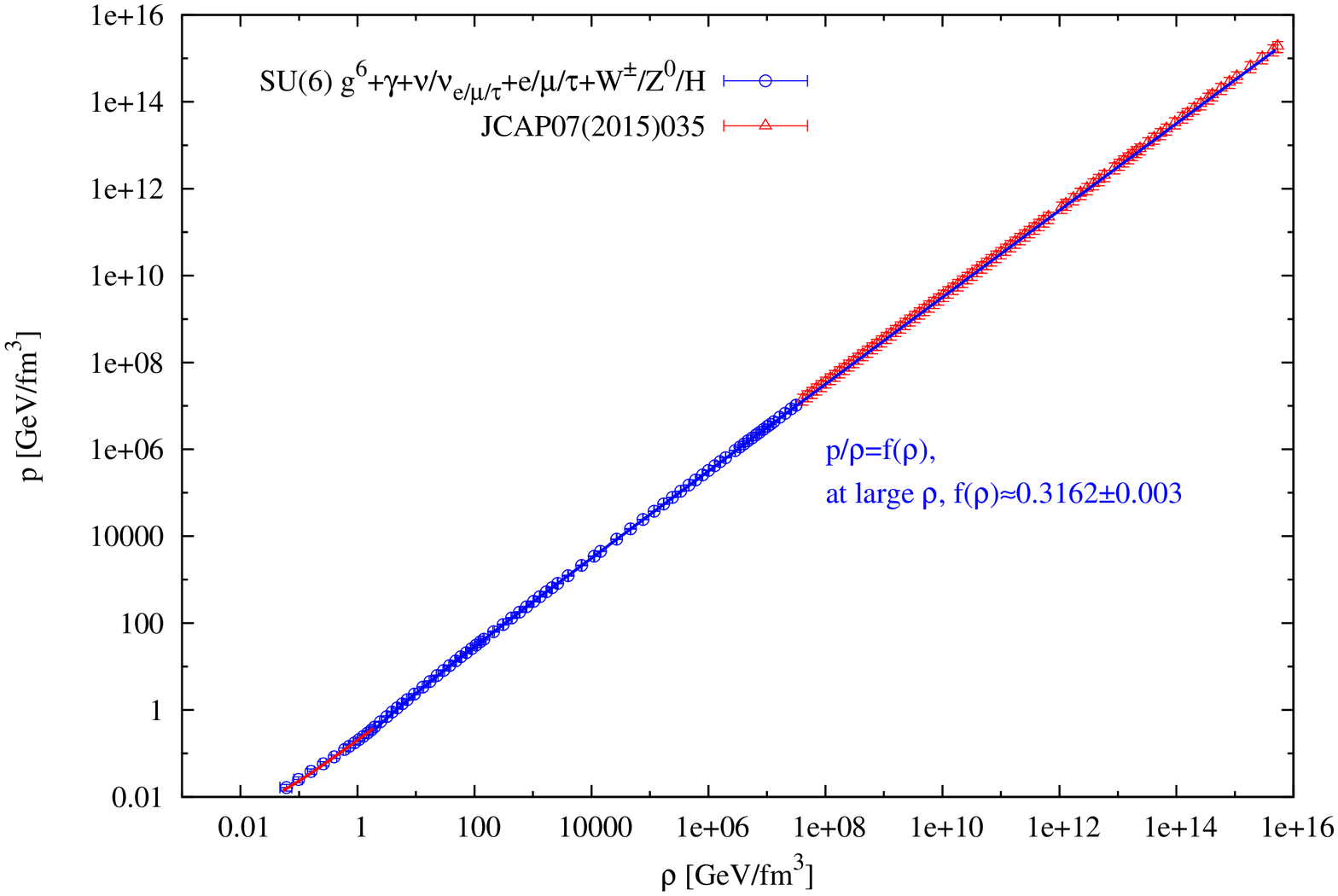}
\caption{Left panel: the thermodynamic pressure calculated in $\mathtt{SU(5)}$ non--perturbative and perturbative lattice QCD in dependence on the corresponding energy density (red and green symbols) \cite{Borsanyi:2016ksw} is confronted to the same calculations extended to other dof; $\gamma$, neutrinos, leptons, EW, and Higgs bosons (blue symbols). Right panel: the extended $\mathtt{SU(5)}$ lattice QCD calculations are combined with $\mathtt{SU(6)g^6}$ non--perturnative calculations \cite{Laine:2015kra,DOnofrio:2015gop}.}
\label{fig:xx2}
}
\end{figure}

Left panel of Fig. \ref{fig:xx2} shows the EoS deduced from $\mathtt{SU(5)}$ non--perturbative and perturbative lattice QCD simulations (red and green symbols) \cite{Borsanyi:2016ksw}. The thermodynamic contributions from $\gamma$, $\nu$, $e$, $\mu$, $\tau$, $\nu_e$, $\nu_{\mu}$, $\nu_{\tau}$, $W^{\pm}$, and the Higgs boson $H$ are summed up with the $\mathtt{SU(5)}$ results (blue symbols). For each type of the additional particles, a partition function is constructed, from which the temperature dependence of the different thermodynamic quantities, such as the thermodynamic pressure and the energy density, can straightforwardly be derived. It is obvious that the additional dof considerably contribute to both thermodynamic quantities. The quantitative contributions are considrable, for example, within the hadronic phase, the proportionality constant in $p\propto\rho$ increases with the additive dof, i.e., increasing speed of sound squared \cite{NasserTawfik:2012jpa}. In quark--gluon plasma and electroweak phases, more structures are added to the one corresponding EoS based on $\mathtt{SU(5)}$ lattice QCD simulations. 

The fitting functions illustrated in the bottom panel of Fig. \ref{fig:xx2} are
\bea
\mathtt{HP:} \qquad \qquad p &=& \alpha_1 + \alpha_2 \rho, \label{eq:eos21} \\
\mathtt{QGP/EW:} \qquad p &=& \beta_1 + \beta_2 \rho + \beta_3 \rho^{d_2}, \label{eq:eos22} \\
\mathtt{Asymp.:} \qquad p  &=& \gamma \rho, \label{eq:eos23}
\eea
where $\alpha_1 = 0.0034 \pm 0.0023$, $\alpha_2  = 0.1991 \pm 0.0022$, 
$\beta_1 = 0.0484 \pm 0.0164$, $\beta_2 = 0.3162 \pm 0.0031$, $\beta_3 = -0.21 \pm 0.014$, and $\gamma = 0.3162 \pm 0.003$.

\subsection{Bulk Viscosity for Strong and Electroweak Matter}

It was concluded that the shear viscosity normalized to the entropy density for perturbative gauge QCD likely approaches the lower bound of Anti--de Sitter/Conformal Field Theory (AdS/CFT) \cite{Kovtun:2004de}. A non--perturbative estimation for viscous coefficients, at temperatures several times the QCD scale, has been reported in ref. \cite{Sakai:2007cm,PhysRevD.98.054515}. This was possible through accumulating a large amount of configurations for the Green function expressed in the Matsubara frequencies and implemented on isotropic $24^2 \times 8$ and $16^2 \times 8$ lattices. The viscous coefficients are determined as slopes of the spectral functions, at vanishing Matsubara frequency. A recent estimation for the temperature dependence of the bulk viscosity of SU($3$) gluodynamics was possible with $48^3 \times 16$ lattice QCD simulations \cite{PhysRevD.98.054515}. Another estimation is based the retarded Green function defined by the Kramers--Kronkig relation in terms of different thermodynamic quantities \cite{Kharzeev:2007wb,Karsch:2007jc}. In ref. \cite{Karsch:2007jc}, it was taken into consideration that the bulk viscosity measures the violation of the conformal invariance. This allowed to conclude that QCD at classical level is conformally invariant. 
\bea
\zeta &=& \frac{1}{9 \omega_0} \Big[T\, s \left(\frac{\partial \rho}{\partial p} -3\right) - 4(\rho - 3p)  \qquad\quad\;\,\mathtt{thermal}\;\;\mathtt{parts} \nonumber \\
&+&  \left(T\frac{\partial}{\partial T} -2 \right) \langle\bar{q}q\rangle(T) + g_g\, G^2(T)  \qquad\quad \mathtt{thermal\;q\;\&\;g\;condensates} \nonumber \\
&+& g_f\left(m_{\pi}^2 f_{\pi}^2 + m_{K}^2 f_{K}^2+ m_D^2 f_D^2 + \cdots\right)\Big].  \;\;\;\mathtt {vacuum\;q\;\&\;g\;condensates} \label{eq:zeta1}
\eea 
where $g_g$ ($g_f$) are the degeneracy factors for gluons (quarks). The spin polarization multiplied is $g_g=16$. The color degrees of freedom reads $N_c^2-1$, with $N_c$ is the color quantum number. $g_f=12\, n_f$ with $n_f$ are the degrees of freedom of the quark flavors. $m_D$ ($f_D$) are mass (decay constant of $D$-meson). The scale parameter $\omega_0$ determines the applicability of the parturbation theory. $\zeta$ is obtained by using the frequency limit of the spectral density, at vanishing spatial momentum \cite{Karsch:2007jc,NoronhaHostler:2008ju} and by implementing the various thermodynamic quantities are detailed in Eq. (\ref{eq:zeta1}).

\begin{figure}[htb]
\centering{
\includegraphics[width=6cm,angle=-90]{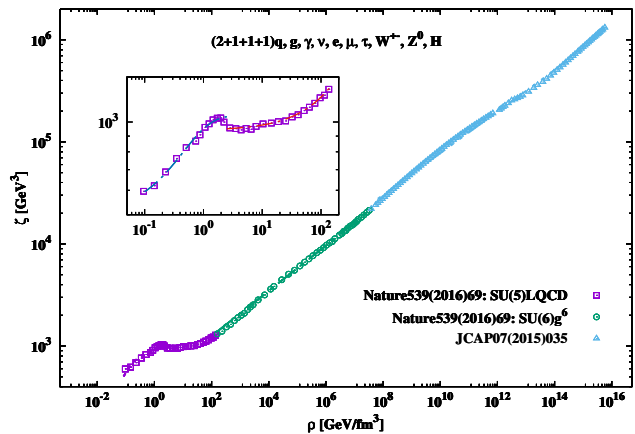}
\caption{\footnotesize The bulk viscosity $\zeta$ in dependence of the energy density $\rho$. Both quantities are calculated, at vanishing baryon--chemical potential and given in physical units. The inside box magnifies the region, at low temperature, where $\zeta(\rho)$ is non--monotonic.}
\label{fig:zeta}}
\end{figure}

By combining the gluons and ($2+1+1+1$) quark contributions and that of the gauge bosons; the photons, $W^{\pm}$, and $Z^0$, of the charged leptons; neutrino, electron, muon, and tau, and of the Higgs bosons; scalar Higgs particle \cite{Tawfik:2019jsa}, the bulk viscosity is shown in Fig. \ref{fig:zeta}. We conclude that the almost entire SM contributions are very significant. The missing SM--contributions are vacuum and thermal bottom quark condensate, the entire gravitational sector, neutral leptons, and top quark. $\zeta$ almost linearly increases with increasing energy--density, so that following three regions of parameterizations (curves) can be distinguished
\bea
\mathtt{Hadron-QGP:} &&  \zeta= a_1+a_2 \rho+a_3 \rho^{a_4}, \label{eq:qcdpt2} \\
\mathtt{QCD:}   && \zeta= b_1 +b_2 \rho^{b_3}, \label{eq:qcdew2} \\
\mathtt{EW:}  && \zeta= c_1 +c_2 \rho^{c_3}. \label{eq:eqpt2}
\eea
For Hadron--QCD: $a_1=-9.336\pm 4.152$, $a_2=0.232\pm 0.003$, $a_3=11.962\pm 4.172$, and $a_4=0.087\pm 0.029$.
For QCD: $b_1= 8.042\pm 0.056$, $b_2= 0.301\pm 0.002$, and $b_3= 0.945\pm 0.0001$.
For EW: $c_1= 0.350\pm 0.065$,  $c_2= 10.019 \pm 0.934$, and $c_3= 0.929 \pm 8.898\times 10^{-5}$.

\subsection{Relaxation Time for Strong and Electroweak Matter}

\begin{figure}[htb]
\centering{
\includegraphics[width=6cm,angle=-90]{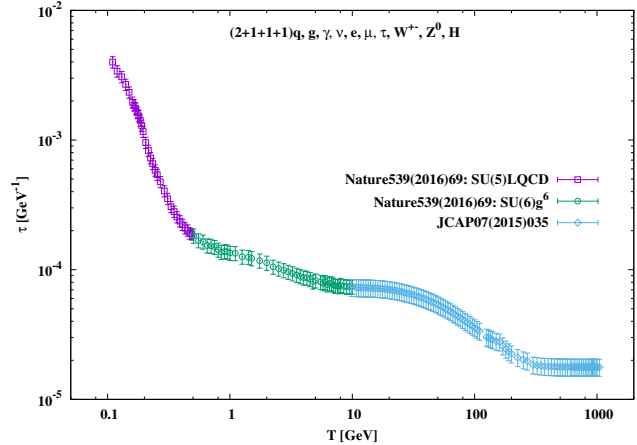}
\caption{\footnotesize At vanishing baryon--chemical potential, the relaxation time $\tau$ is presented in dependence on the temperature $T$. The results are based on non--perturbative and perturbative lattice QCD simulations, in which contributions from the quark and gluon condensates and the thermodynamic quantities of gauge bosons, charged leptons, and Higgs bosons (left symbols) are integrated. 
\label{fig:tau}}
}
\end{figure}

The relaxation time $\tau_f$, which involves complicated collision integrals, can be determined as the mean collision time and averages of thermodynamic quantities \cite{Tawfik:2016edq,Tawfik:2011sh,Tawfik:2010bm}
\bea
\tau_f(T) &=& \frac{1}{n_f(T)\, \langle v(T)\rangle \sigma(T)}, \label{eq:tau1}
\eea
where $\sigma$ is cross section, $\langle v(T)\rangle$ stands for mean relative velocity, and $n_f(T)$ represents the number density. Approaches for $\langle v(T)\rangle$ and $\sigma$ have been discussed in ref. \cite{Tawfik:2010bm}. The temperature dependence of the relaxation time $\tau_f(T)$ plays an essential role in bulk viscosity $\zeta(T)$. 

Figure \ref{fig:tau} presents $\tau(T)$ deduced from the non--perturbative and perturbative QCD simulations \cite{Borsanyi:2016ksw,Laine:2015kra,DOnofrio:2015gop,Tawfik:2019jsa} (bottom symbols). Contributions from the quark and gluon condensates and the thermodynamic quantities of the gauge bosons, charged leptons, and Higgs bosons are added (left symbols). We notice that $\tau$ steadily decreases with increasing $T$. In the different phases, there are different rates of decreasing $\tau$.

\section{Conclusions}
\label{sec:cncl}

Most reliable non--perturbative lattice QCD simulations and perturbative calculations with as much as possible quark flavors with physical masses are phenomenologically combined. The thermodynamic contributions of photons, charged neutrinos, leptons, electroweak particles ($W^{\pm}$ and $Z^0$ bosons), and the scalar Higgs bosons are also integrated in. This makes the present study pioneering in a) simultaneous accessing hadron, quark--gluon plasma and electroweak epochs of the early Universe and b) including almost all degrees of freedom of the standard model. With this regard, we emphasize that the only missing SM dof are vacuum and thermal bottom quark condensate, the entire gravitational sector, neutral leptons, and top quark flavor. 

We have introduced various thermodynamic quantities, including pressure, energy density, bulk viscosity, and relaxation time, at vanishing net--baryon cosmic matter and temperatures up to the TeV--scale. The main result is that recent non--perturbative lattice QCD simulations and perturbative calculations jointly lead to EoS covering a wide range of temperatures. As never done before, this study makes introduces a framework to combine the standard model of the elementary particles and the standard model for cosmology.


\bibliographystyle{ws-procs961x669}
\bibliography{ATawfikRefsMG16-Talk1}

\begin{thebibliography}{10}

\bibitem{misner1968isotropy}
C.~W. Misner, The isotropy of the universe, {\em The Astrophysical Journal}
  {\bf 151}, p. 431  (1968).

\bibitem{Zeldovich:1983cr}
{\relax Ya}.~B. Zeldovich and I.~D. Novikov, {\em {RELATIVISTIC ASTROPHYSICS.
  VOL. 2. THE STRUCTURE AND EVOLUTION OF THE UNIVERSE}} 1983.

\bibitem{husdal2016viscosity}
L.~Husdal, Viscosity in a lepton-photon universe, {\em Astrophysics and Space
  Science} {\bf 361}, 1  (2016).

\bibitem{Tawfik:2004sw}
A.~Tawfik, {QCD phase diagram: A Comparison of lattice and hadron resonance gas
  model calculations}, {\em Phys. Rev. D} {\bf 71}, p. 054502  (2005).

\bibitem{Tawfik:2010pm}
A.~Tawfik, M.~Wahba, H.~Mansour and T.~Harko, {Hubble Parameter in QCD Universe
  for finite Bulk Viscosity}, {\em Annalen Phys.} {\bf 522}, 912  (2010).

\bibitem{STAR:2015vvs}
L.~Adamczyk {\em et~al.}, {Probing parton dynamics of QCD matter with $\Omega$
  and $\phi$ production}, {\em Phys. Rev. C} {\bf 93}, p. 021903  (2016).

\bibitem{Hakk:2021aav}
E.~A. Hakk, A.~N. Tawfik, A.~Nada and H.~Yassin, {Cosmic Evolution of Viscous
  QCD Epoch in Causal Eckart Frame}, {\em Universe} {\bf 7}, p. 112  (2021).

\bibitem{Tawfik:2019jsa}
A.~N. Tawfik and I.~Mishustin, {Equation of State for Cosmological Matter at
  and beyond QCD and Electroweak Eras}, {\em J. Phys. G} {\bf 46}, p. 125201
  (2019).

\bibitem{Borsanyi:2016ksw}
S.~Borsanyi {\em et~al.}, {Calculation of the axion mass based on
  high-temperature lattice quantum chromodynamics}, {\em Nature} {\bf 539}, 69
  (2016).

\bibitem{Borsanyi:2012zs}
S.~Borsanyi {\em et~al.}, {High-precision scale setting in lattice QCD}, {\em
  JHEP} {\bf 09}, p. 010  (2012).

\bibitem{Cheng:2007wu}
M.~Cheng, {Charm Quarks and the QCD Equation of State}, {\em PoS} {\bf
  LATTICE2007}, p. 173  (2007).

\bibitem{Levkova:2009gq}
L.~Levkova, {Effects of the charm quark on the QCD equation of state}, {\em
  PoS} {\bf LAT2009}, p. 170  (2009).

\bibitem{Burger:2013hia}
F.~Burger, G.~Hotzel, M.~M\"uller-Preussker, E.-M. Ilgenfritz and M.~P.
  Lombardo, {Towards thermodynamics with $N_f=2+1+1$ twisted mass quarks}, {\em
  PoS} {\bf Lattice2013}, p. 153  (2013).

\bibitem{MILC:2013ops}
A.~Bazavov {\em et~al.}, {Update on the 2+1+1 Flavor QCD Equation of State with
  HISQ}, {\em PoS} {\bf LATTICE2013}, p. 154  (2014).

\bibitem{Borsanyi:2011wyg}
S.~Borsanyi, G.~Endrodi, Z.~Fodor, S.~D. Katz, S.~Krieg, C.~Ratti, C.~Schroeder
  and K.~K. Szabo, {The QCD equation of state and the effects of the charm},
  {\em PoS} {\bf LATTICE2011}, p. 201  (2011).

\bibitem{Bellwied:2015lba}
R.~Bellwied, S.~Borsanyi, Z.~Fodor, S.~D. Katz, A.~Pasztor, C.~Ratti and K.~K.
  Szabo, {Fluctuations and correlations in high temperature QCD}, {\em Phys.
  Rev. D} {\bf 92}, p. 114505  (2015).

\bibitem{Andersen:2010wu}
J.~O. Andersen, L.~E. Leganger, M.~Strickland and N.~Su, {NNLO
  hard-thermal-loop thermodynamics for QCD}, {\em Phys. Lett. B} {\bf 696}, 468
   (2011).

\bibitem{Laine:2006cp}
M.~Laine and Y.~Schroder, {Quark mass thresholds in QCD thermodynamics}, {\em
  Phys. Rev. D} {\bf 73}, p. 085009  (2006).

\bibitem{Kajantie:2002wa}
K.~Kajantie, M.~Laine, K.~Rummukainen and Y.~Schroder, {The Pressure of hot QCD
  up to g6 ln(1/g)}, {\em Phys. Rev. D} {\bf 67}, p. 105008  (2003).

\bibitem{Brambilla:2006wp}
N.~Brambilla, X.~Garcia~i Tormo, J.~Soto and A.~Vairo, {The Logarithmic
  contribution to the QCD static energy at N**4 LO}, {\em Phys. Lett. B} {\bf
  647}, 185  (2007).

\bibitem{Laine:2015kra}
M.~Laine and M.~Meyer, {Standard Model thermodynamics across the electroweak
  crossover}, {\em JCAP} {\bf 07}, p. 035  (2015).

\bibitem{Laine:2013raa}
M.~Laine, G.~Nardini and K.~Rummukainen, {First order thermal phase transition
  with 126 GeV Higgs mass}, {\em PoS} {\bf LATTICE2013}, p. 104  (2014).

\bibitem{Linde:1980ts}
A.~D. Linde, {Infrared Problem in Thermodynamics of the Yang-Mills Gas}, {\em
  Phys. Lett. B} {\bf 96}, 289  (1980).

\bibitem{Gross:1980br}
D.~J. Gross, R.~D. Pisarski and L.~G. Yaffe, {QCD and Instantons at Finite
  Temperature}, {\em Rev. Mod. Phys.} {\bf 53}, p.~43  (1981).

\bibitem{Gynther:2005dj}
A.~Gynther and M.~Vepsalainen, {Pressure of the standard model at high
  temperatures}, {\em JHEP} {\bf 01}, p. 060  (2006).

\bibitem{Kajantie:1996mn}
K.~Kajantie, M.~Laine, K.~Rummukainen and M.~E. Shaposhnikov, {Is there a~ hot
  electroweak phase transition at $m_H \gtrsim m_W$?}, {\em Phys. Rev. Lett.}
  {\bf 77}, 2887  (1996).

\bibitem{Aoki:1999fi}
Y.~Aoki, F.~Csikor, Z.~Fodor and A.~Ukawa, {The Endpoint of the first order
  phase transition of the SU(2) gauge Higgs model on a four-dimensional
  isotropic lattice}, {\em Phys. Rev. D} {\bf 60}, p. 013001  (1999).

\bibitem{Brodsky:2002nb}
S.~J. Brodsky, S.~Menke, C.~Merino and J.~Rathsman, {On the behavior of the
  effective QCD coupling alpha(tau)(s) at low scales}, {\em Phys. Rev. D} {\bf
  67}, p. 055008  (2003).

\bibitem{DOnofrio:2015gop}
M.~D'Onofrio and K.~Rummukainen, {Standard model cross-over on the lattice},
  {\em Phys. Rev. D} {\bf 93}, p. 025003  (2016).

\bibitem{NasserTawfik:2012jpa}
A.~Nasser~Tawfik and H.~Magdy, {Hadronic Equation of State and Speed of Sound
  in Thermal and Dense Medium}, {\em Int. J. Mod. Phys. A} {\bf 29}, p. 1450152
   (2014).

\bibitem{Kovtun:2004de}
P.~Kovtun, D.~T. Son and A.~O. Starinets, {Viscosity in strongly interacting
  quantum field theories from black hole physics}, {\em Phys. Rev. Lett.} {\bf
  94}, p. 111601  (2005).

\bibitem{Sakai:2007cm}
S.~Sakai and A.~Nakamura, {Lattice calculation of the QGP viscosities: Present
  results and next project}, {\em PoS} {\bf LATTICE2007}, p. 221  (2007).

\bibitem{PhysRevD.98.054515}
N.~Y. Astrakhantsev, V.~V. Braguta and A.~Y. Kotov, Temperature dependence of
  the bulk viscosity within lattice simulation of $su(3)$ gluodynamics, {\em
  Phys. Rev. D} {\bf 98}, p. 054515 (Sep 2018).

\bibitem{Kharzeev:2007wb}
D.~Kharzeev and K.~Tuchin, {Bulk viscosity of QCD matter near the critical
  temperature}, {\em JHEP} {\bf 09}, p. 093  (2008).

\bibitem{Karsch:2007jc}
F.~Karsch, D.~Kharzeev and K.~Tuchin, {Universal properties of bulk viscosity
  near the QCD phase transition}, {\em Phys. Lett.} {\bf B663}, 217  (2008).

\bibitem{NoronhaHostler:2008ju}
J.~Noronha-Hostler, J.~Noronha and C.~Greiner, {Transport Coefficients of
  Hadronic Matter near T(c)}, {\em Phys. Rev. Lett.} {\bf 103}, p. 172302
  (2009).

\bibitem{Tawfik:2016edq}
A.~N. Tawfik, A.~M. Diab and M.~T. Hussein, {SU(3) Polyakov linear-sigma model:
  Conductivity and viscous properties of QCD matter in thermal medium}, {\em
  Int. J. Mod. Phys.} {\bf A31}, p. 1650175  (2016).

\bibitem{Tawfik:2011sh}
A.~Tawfik and T.~Harko, {Quark-Hadron Phase Transitions in Viscous Early
  Universe}, {\em Phys. Rev.} {\bf D85}, p. 084032  (2012).

\bibitem{Tawfik:2010bm}
A.~Tawfik, M.~Wahba, H.~Mansour and T.~Harko, {Viscous Quark-Gluon Plasma in
  the Early Universe}, {\em Annalen Phys.} {\bf 523}, 194  (2011).

\end{thebibliography}


\end{document}